\documentclass[fleqn,usenatbib]{mnras}

\usepackage{newtxtext,newtxmath}

\usepackage[T1]{fontenc}

\DeclareRobustCommand{\VAN}[3]{#2}
\let\VANthebibliography\thebibliography
\def\thebibliography{\DeclareRobustCommand{\VAN}[3]{##3}\VANthebibliography}

\usepackage{graphicx}	
\usepackage{amsmath}	

\title[]{Spectroscopic confirmation of a gravitationally lensed Lyman break galaxy at $z$$_{\rm[CII]}$= 6.827 using NOEMA}

\author[S. J. Molyneux et al.]{
S. J. Molyneux$^{1,2}$,\thanks{E-mail: s.j.molyneux@2019.ljmu.ac.uk} R. Smit$^{1}$, D. Schaerer$^{3,4}$, R. J. Bouwens$^{5}$, L. Bradley$^{6}$, J. A. Hodge$^{5}$, S. N. Longmore$^{1}$, \newauthor S. Schouws$^{5}$, P. van der Werf$^{5}$, A. Zitrin$^{7}$, S. Phillips$^{1}$
\\
$^{1}$Astrophysics Research Institute, Liverpool John Moores University, 146 Brownlow Hill, Liverpool L3 5RF, UK \\
$^{2}$ European Southern Observatory, Karl-Schwarzschild-str. 2, 85748 Garching, Germany\\
$^3$ Observatoire de Gen\`{e}ve, Universit\'{e} de Gen\`{e}ve, 51 Ch. des Mail-lettes, 1290 Versoix, Switzerland \\
$^4$ CNRS, IRAP, 14 avenue E. Belin, 31400 Toulouse, France\\
$^5$ Leiden Observatory, Leiden University, NL-2300 RA Leiden, Netherlands\\
$^6$ Space Telescope Science Institute, Baltimore, MD 21218, USA\\
$^7$ Physics Department,  Ben-Gurion University of the Negev, P.O. Box 653, Be'er-Sheva 8410501, Israel
}

\date{Accepted XXX. Received YYY; in original form ZZZ}

\pubyear{2021}

\begin{document}
\label{firstpage}
\pagerange{\pageref{firstpage}--\pageref{lastpage}}
\maketitle

\begin{abstract}
We present the spectroscopic confirmation of the brightest known gravitationally lensed Lyman break galaxy in the Epoch of Reionisation, A1703-zD1, through the detection of [C~{\sc ii}]158$\mu$m at a redshift of $z$ = 6.8269 $\pm$ 0.0004. This source was selected behind the strong lensing cluster Abell 1703, with an intrinsic $L_{\rm UV} \sim L^*_{\rm z=7}$ luminosity and a very blue \emph{Spitzer}/IRAC [3.6]--[4.5] colour, implying high equivalent width line emission of [O~{\sc iii}]+H$\beta$. [C~{\sc ii}] is reliably detected at 6.1$\sigma$ co-spatial with the rest-frame UV counterpart, showing similar spatial extent. Correcting for the lensing magnification, the [C~{\sc ii}] luminosity in A1703-zD1 is broadly consistent with the local $L_{\rm[CII]}$ -- SFR relation. We find a clear velocity gradient of 103 $\pm$ 22 km~$\rm s^{-1}$ across the source which possibly indicates rotation or an ongoing merger. We furthermore present spectral scans with no detected [C~{\sc ii}] above 4.6$\sigma$ in two unlensed Lyman break galaxies in the EGS-CANDELS field at $z$ $\sim$ 6.6 -- 6.9. This is the first time that NOEMA has been successfully used to observe [C~{\sc ii}] in a `normal' star-forming galaxy at $z>6$, and our results demonstrate its capability to complement ALMA in confirming galaxies in the Epoch of Reionisation.

\end{abstract}




\section{Introduction}
\label{sec:Introdcution}

In the past decade hundreds of galaxies have been identified in the Epoch of Reionisation (EoR), selected from their rest-frame UV light, using Hubble Space Telescope (\emph{HST}) and ground-based optical/NIR observatories \citep[][for a review and references therein]{Stark16}. However, only a fraction of these sources have spectroscopic redshift determinations and we have a limited understanding of their physical properties. One reason for this is the difficulty in obtaining Lyman-$\alpha$ observations at such high redshifts, due to its absorption by neutral hydrogen in the intergalactic and interstellar medium (with detected Ly$\alpha$ emitters possibly residing in early ionised bubbles; e.g. \citealt{Jung20,Endsley21}).

In recent years ALMA has transformed this field by confirming the redshifts of galaxies out to redshift $z$ = 9 \citep[e.g.,][]{Smit18,Hashimoto18,Tamura19, Hodge20, Bouwens21} and providing the first view of their dust obscured star-formation \citep[e.g.,][]{Watson15, Laporte17,Bowler18,Schouws21}, the kinematics of these sources \citep[e.g.,][]{Smit18, Hashimoto19,Fujimoto19, Ginolfi19}, the cool gas traced by [C~{\sc ii}] and highly ionised gas traced by [O~{\sc iii}] \citep[e.g.,][]{Maiolino15, Inoue16, Carniani17, Laporte17, Hashimoto19, Tamura19, Harikane19, Bakx20}. Uncovering the physical properties of these primordial systems is fundamental to understanding the evolution of the first generation of galaxies and their role in Cosmic Reionisation.

In particular, ALMA has demonstrated its ability to detect [C~{\sc ii}] in UV bright galaxies which are intrinsically bright ($\sim$2~--~3~$\times$ $L^\ast$ at $z$ = 7) at $z\gtrsim5$ \citep{Capak15,Willott15,Smit18,Matthee19,Bethermin20,Bouwens21}. However, much less is known about galaxies at these high redshifts which are more representative of the galaxy population as a whole ($L \leq L^\ast_{\rm z=7}$). Our best insights into this fainter population are rare, apparently bright sources, which have been gravitationally lensed \citep{Knudsen16,Bradac17,Fujimoto21,Laporte21}. Despite the capability of ALMA to probe galaxies in the EoR, it has limited sky coverage in the Northern Hemisphere and so interesting targets might be missed.


Recently, the Plateau de Bure Interferometer (PdBI) has been upgraded to the Northern Extended Millimeter Array (NOEMA). The upgrades have increased the number of antennae from 6 to 10 (with 2 more planned to eventually increase the total number to 12) providing more collecting power and therefore increased sensitivity to observe these fainter sources. A new correlator has also been installed (upgrading from Widex to PolyFix \citealt{Schuster18}), which can cover a larger frequency range in one setup, enabling faster line scans. These upgrades arguably make NOEMA the most powerful interferometer in the Northern Hemisphere, and therefore might play an important role in observing [C~{\sc ii}] in galaxies at $z$ > 6.

Here we report on a line search for [C~{\sc ii}]158$\mu$m with NOEMA, targeting 3 galaxies in the Northern Hemisphere at $z\sim6.6-6.9$. The targets have been selected from a larger sample of Lyman Break Galaxies, with high-precision photometric redshifts \citep{Smit14,Smit15}. A1703-zD1 is the standout target of our sample, lying behind the strong lensing cluster Abell 1703, with a magnification of $\sim$9 \citep{Bradley12}. Due to its exceptional observed brightness, it has been targeted many times with previous observations attempting to observe Ly$\alpha$, C~{\sc iv} and C~{\sc iii} with the Keck Observatory \citep{Schenker12, Stark15b, Mainali18} and a previous attempt with PdBI \citep{Schaerer15} to observe [C~{\sc ii}], however, these observations did not result in a significant detection.

In this paper we present the successful spectroscopic confirmation of A1703-zD1 and constraints on the properties of the other two sources based on their non-detections. The plan of the paper is as follows. In Section~\ref{sec:Sample} we describe the sample selection and observations. In Section~\ref{sec:LineScans} we discuss the method for line scanning and the findings from our scans. In Section~\ref{sec:SourceProperties} we provide the properties of the sources and comparisons to the literature and in Section~\ref{sec:Conclusions} we give our summary and conclusions. In this paper we adopt a Kroupa IMF \citep{Kroupa01}. We adopt $H_0=70$\,km~$\rm s^{-1}$\,Mpc$^{-1}$, $\Omega_M=0.3$, and $\Omega_\Lambda=0.7$ throughout. Magnitudes are quoted in the AB system \citep{Oke83}.

\section{Sample Selection and Observations}
\label{sec:Sample}

\subsection{Target selection and properties}
\label{sec:targets}

We obtained NOEMA observations of three sources, A1703-zD1 \citep{Bradley12, Smit14}, EGS-5711424617 and EGS-1952445714 \citep[hereafter EGS-5711 and EGS-1952 respectively;][]{Smit15}.
These galaxies were initially selected with the Lyman break technique as \emph{HST}/F814W drop-out galaxies and subsequently identified as sources with blue \emph{Spitzer}/IRAC [3.6]--[4.5] colours, implying high equivalent width [O~{\sc iii}]+H$\beta$ emission \citep{Smit14,Smit15}. The spectral energy distribution (SED) for each galaxy is shown in Appendix~\ref{app:A}.

We selected these targets due to their observed brightness ($m_{\rm UV}$ $\sim$ 24 -- 25), but also as they are representative of `normal' star-forming galaxies (SFR < 100 M$_\odot$ yr$^{-1}$) at $z \sim$ 7, with intrinsic
(corrected for lensing magnification, A1703-zD1 is magnified $\sim$9$\times$)  
UV SFRs of 5 -- 38 M$_\odot$ yr$^{-1}$. A list of some of the basic properties for all three galaxies are shown in Table~\ref{tab:Sample}.
These same selection criteria were recently used to successfully confirm galaxies with ALMA at $z_{\rm [CII]} = 6.808-6.854$ \citep{Smit18}. 

The inferred emission lines in the \emph{Spitzer} observations reduce the probability range for the redshift such that observations can be carried out using one NOEMA setup.
For both A1703-zD1 and EGS-1952 the photometric redshifts of $6.7_{-0.1}^{+0.2}$ and $6.75_{-0.10}^{+0.11}$, respectively, are within this range. However, we note that EGS-5711 has $z_{\rm phot}$ = $6.47_{-0.10}^{+0.11}$, outside of the colour selection range due to a tentative detection in the \emph{HST}/F814W band. For the extreme emission line sources from \citep{Smit15}, systematic changes in the estimated photometric redshift ($\Delta z\sim 1$) are found when making changes to the input template set (in particular the strength of the emission line equivalent widths of the templates) used to fit the SEDs. As a result of these systematic uncertainties, we rely on the Spitzer/IRAC colour selection from \citet{Smit15} to identify sources most likely in the redshift range $z\sim6.6-6.9$, which includes some cases (like EGS-5711) where the $z_{\rm phot}$ is not within this range.

For A1703-zD1 we use a magnification value of $9.0_{-4.4}^{+0.9}$ taken from \citet{Bradley12} calculated using the model described in \citealt{Zitrin10}. The magnification error represents the extreme value obtained from the minimum and maximum magnifications obtained within 0.5 arcsec of the target and assuming $\Delta$z $\pm$ 1.0 for the source redshift. To obtain a better handle on the systematic uncertainties in the lensing magnification we ran two more trial models, one using a revised version of the Light Traces Mass (LTM) technique and one fully parametric model. These models suggest a magnification in agreement with that of \cite{Bradley12} within the uncertainties, though towards the lower end ($\mu$ $\sim$ 4 -- 5). Critically, the weight of the nearby cluster galaxy (seen in the bottom of Figure~\ref{fig:A1703Contour}) is fixed according to the scaling relations, but in reality has a significant uncertainty that affects the magnification value. We therefore adopt the published magnification by \citet{Bradley12} ($\mu$=$9.0_{-4.4}^{+0.9}$), as the uncertainties are broad enough to include the new magnifications from the trial models.

Throughout the paper, we report measured quantities (i.e. [C~{\sc ii}] flux) as observed in the image plane, without a magnification correction, whereas derived physical quantities (i.e. $L_{\rm[CII]}$, SFR$_{\rm[CII]}$ and physical size) are corrected for the adopted lensing magnification.

\subsection{NOEMA observations and data reduction}
\label{sec:observations}

We obtained 1.2 mm observations using NOEMA in its most compact 10D configuration, with a single setup for each of the three sources, A1703-zD1, EGS-5711 and EGS-1952, approved in program W18FC. 

Observations of A1703-zD1 were taken on 21 March 2019, with 4.1 hours on source, and covering the frequency range 241.45 -- 249.08 GHz (upper side band, USB) and 225.97 -- 233.60 GHz (lower side band, LSB), corresponding to a redshift range 6.63 -- 6.87 and 7.14 -- 7.41 respectively. 
The USB frequency range partly overlaps with PdBI observations taken in 2013 by \cite{Schaerer15} and we combine the NOEMA and PdBI data in the $UV$-plane to obtain maximum depth over the redshift range of 6.80 -- 6.88. 

We observed EGS-5711 on 16 April, 30 April 2019 and 02 May with 7.1 hours on source in total. EGS-1952 was observed on 08 and 09 April 2019 as well as on 02 and 03 May 2019 with 8.7 hours on source in total. Observations for both of these targets covered the same frequency and redshift range. LSB coverage was 240.4 -- 248.3 GHz, corresponding to redshift of 6.66 -- 6.91 and USB coverage was 255.8 -- 263.7 GHz, corresponding to redshift of 6.21 -- 6.43. The coverage described here is shown alongside the SED in Appendix~\ref{app:A}.

All of the data were reduced using the GILDAS software. From the NOEMA data of A1703-zD1 alone, we obtain a datacube with a median rms of 0.87mJy in a 40 km~$\rm s^{-1}$ channel, with a beam size of 1.52 $\times$ 1.28 arcsec. After merging with PdBI data \citep{Schaerer15} we obtain a datacube with a median rms of 0.83mJy in a 40 km~$\rm s^{-1}$ channel, with a beam size of 1.57 $\times$ 1.31 arcsec ($\sim$ 2 times lower sensitivity than ALMA observations in \citealt{Smit18}). We then tapered the datacube for A1703-zD1 such that the beam size matched the spatial extent of the source in the HST imaging using a 50 metre taper at a 170 degree angle, in order to obtain a more accurate measurement of the total flux. We also cleaned the data with 100 iterations before imaging the datacube, including a threshold of 0.85mJy (2 $\times$ rms) to reduce the effects of contamination from a strong CO(3-2) emission line signal observed from a serendipitous source. This leads to a median rms in the datacube of 0.73mJy in a 40 km~$\rm s^{-1}$ channel, with a beam size of 2.88 $\times$ 1.48 arcsec.

We imaged the two EGS datacubes with natural weighting for optimal point source sensitivity, resulting in a beam size of 1.69 $\times$ 1.41 arcsec and 1.63 $\times$ 1.37 arcsec for EGS-5711 and  EGS-1952 respectively. Our data reached a typical rms of 0.34 and 0.41 mJy in a 40 km~$\rm s^{-1}$ channel respectively. We do not apply any tapering as our beam sizes are expected to cover any observable [C~{\sc ii}] emission from these compact targets \citep[][]{Carniani20}.

We also produced continuum images using GILDAS for all three targets to provide constraints on the dust obscured star formation and on the dust content itself. For all three targets we merged the USB and LSB before making the continuum images. In the case of A1703-zD1 we removed the frequency range in which [C~{\sc ii}] was detected in the corresponding side band. We find no continuum detections above 3$\sigma$ within 1 arcsec of each source and therefore provide upper limits for the $L_{\rm IR}$ and SFR$_{\rm IR}$, presented in Table~\ref{tab:GalProperties}. These upper limits are derived by taking 3 $\times$ rms from the continuum image. Discussion about the significance of these continuum measurements can be found in Section~\ref{sec:IR}.

\begin{figure}
    \centering
    \includegraphics[width=0.5\textwidth]{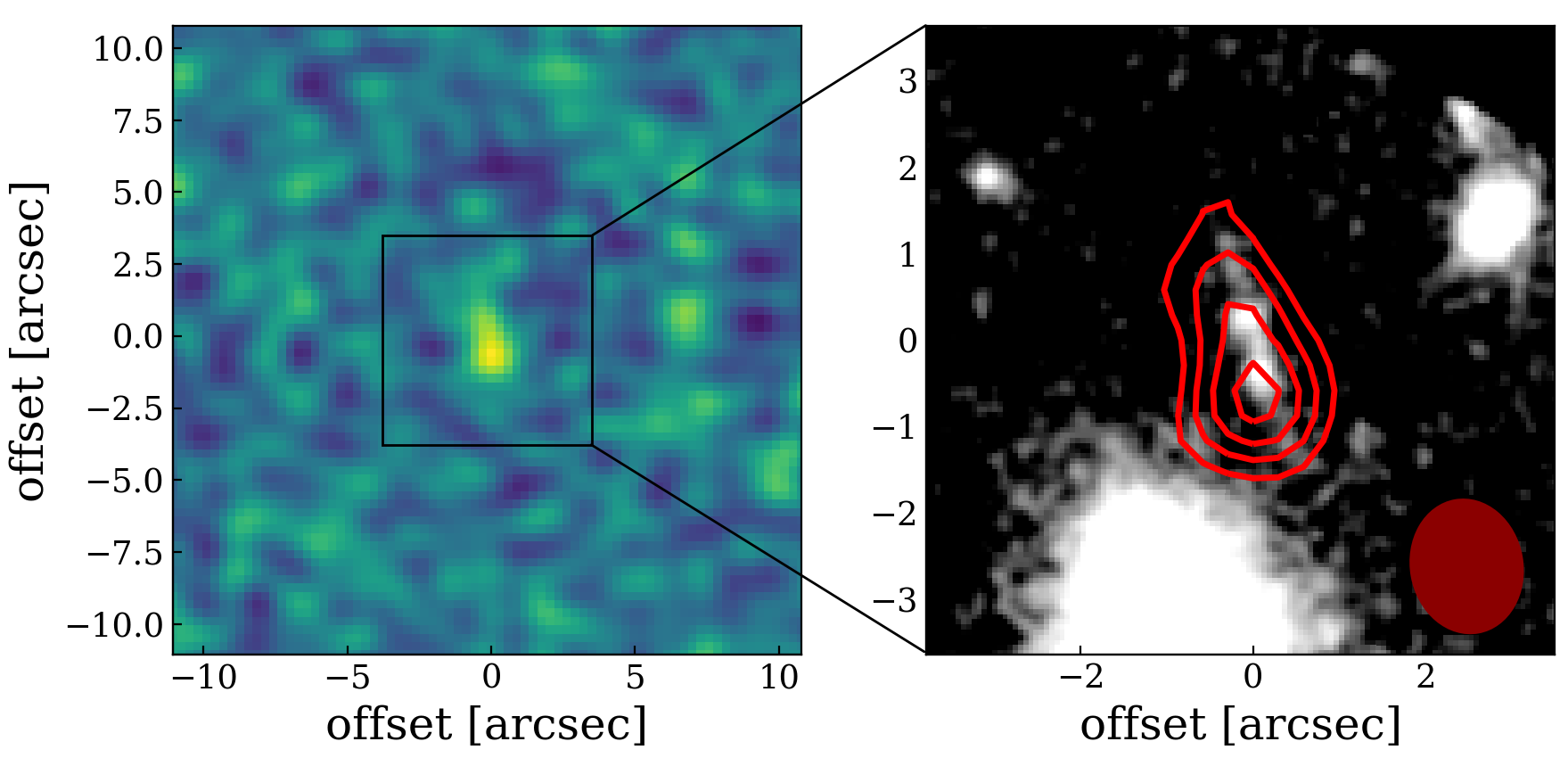}
    \includegraphics[width=0.5\textwidth]{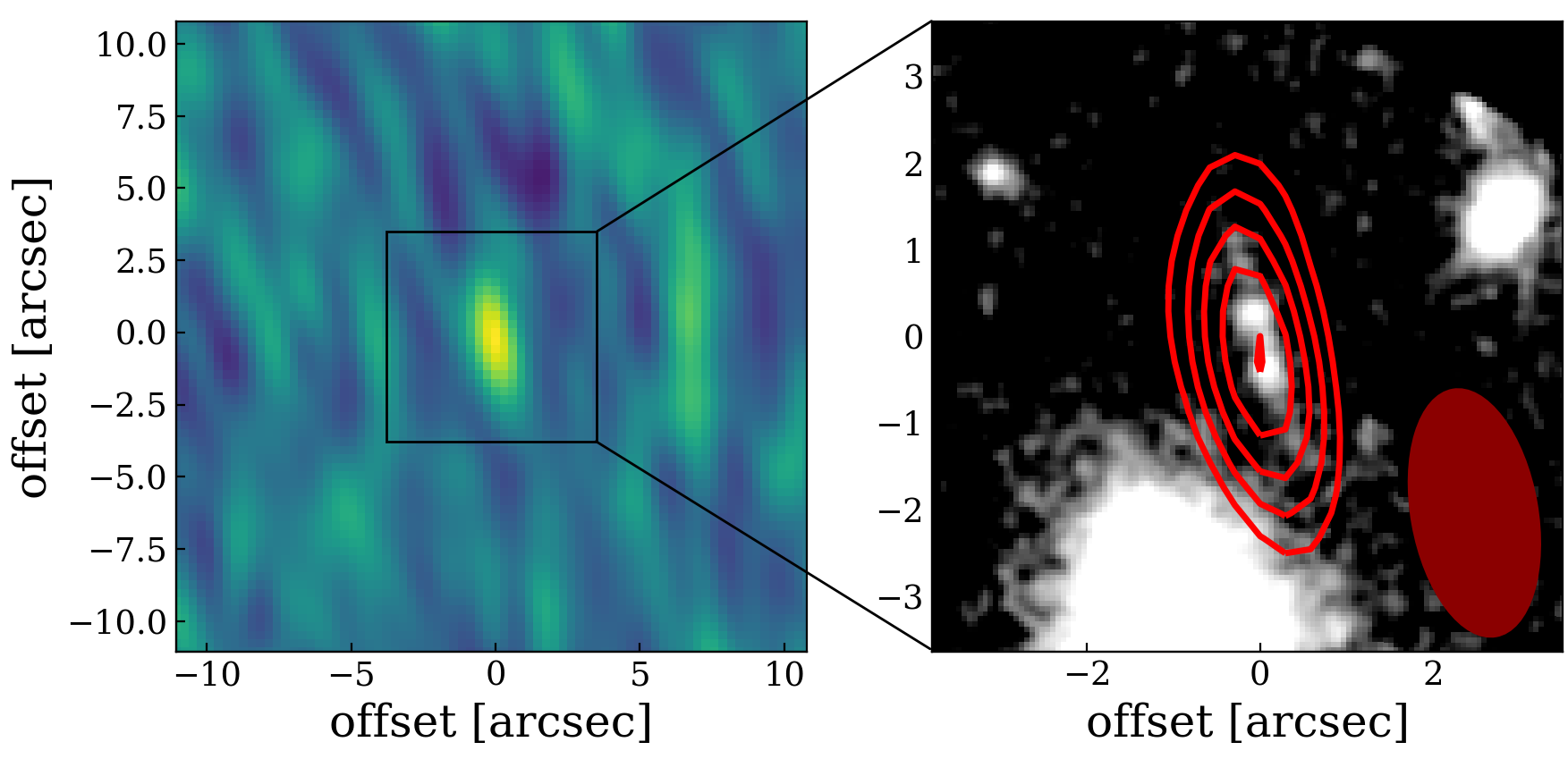}
    \caption{The detection of [C~{\sc ii}] at $z$ = 6.827 in A1703-zD1. The left panels show the NOEMA+PdBI data collapsed over the frequency range 242.76-242.86 GHz for the untapered (\textbf{top panels}) and tapered (\textbf{bottom panels}) imaging. The right panels show HST $H_{160}$ imaging (grey-scale image) overlaid with the 2, 3, 4, 5 and 6 $\sigma$ contours of the [C~{\sc ii}] narrowband (red contours). The filled ellipses in the bottom right corner indicate the beam sizes (1.57 $\times$ 1.31 arcsec and 2.88 $\times$ 1.48 arcsec). }
    \label{fig:A1703Contour}
\end{figure}

We correct for any offset in the HST astrometry by identifying the closest star to the target galaxy in the Gaia Data Release 3 (DR3). For EGS-5711 and EGS-1952 the observations were taken $\sim$ 4 years apart and we correct for the proper motion of the star during that timeframe, before shifting our HST image to the Gaia reference star. Unfortunately for A1703-zD1, none of the stars present in the HST image had proper motion data from Gaia DR3 as a result of being too faint (G magnitudes > 21). Given that the HST imaging for Abell 1703 was taken in 2004 ($>$10 years apart from Gaia), and assuming a similar proper motion to the stars in the EGS field, we expect a $\sim$ 0.3 arcsec offset from the reported Gaia DR3 position due to proper motion. This angular distance is similar to the typical spatial offset found for the stars in the Abell 1703 field between the HST imaging and Gaia DR3 catalogue. As a result, no improvement in the astrometry of A1703-zD1 could be obtained and we relied on the original HST imaging \citep[see][]{Bradley12}, but note that a maximum 0.3 arcsec astrometrical offset between the HST and NOEMA data is possible. However, we point out that the morphology and size of the [C~{\sc ii}] emission for A1703-zD1 matches well to the HST image, as shown in Figure~\ref{fig:A1703Contour}, so any real offset is perhaps likely to be small. Further, the detection of a serendipitous source with much higher S/N is also present and is seen to be co-spatial with the host galaxy as shown in Figure~\ref{fig:A1703serenim}.

\begin{table}
\begin{tabular}{|c|c|c|c|}
\hline
Target ID & A1703-zD1 & EGS-5711424617 & EGS-1952445714\\ 
\hline    
RA & 13:14:59.418 & 14:19:57.114 & 14:19:19.524\\
DEC & +51:50:00.84 & +52:52:46.17 & +52:44:57.14 \\
$z_{\rm phot}$ & $6.7_{-0.1}^{+0.2}$ $^a$ & $6.47_{-0.10}^{+0.11}$ $^b$  & $6.75_{-0.10}^{+0.11}$ $^b$  \\
$ m_{H_{160}}$ &  24.0 $\pm$ 0.1 $^a$ & 25.1 $\pm$ 0.1$^b$ & 25.3 $\pm$ 0.1$^b$\\
$M_{\rm UV}$ & $-$20.6 $\pm$ 0.5$^c$ & $-$21.77$\pm$0.10 & $-$21.64$\pm$0.10 \\
$\rm \beta_{UV}$  & $-$1.56$\pm$0.32 & $-$2.18$\pm$0.36 & $-$2.36$\pm$0.43 \\
$\mu$ & $9.0_{-4.4}^{+0.9}$ $^a$ & -- & -- \\
\hline
\end{tabular}
\caption{
Table of source properties based on previous observations. $^a$Values from \citet{Bradley12}. $^b$Values from \citet{Smit15}. $^c$Values from \citet{Smit14}.}
\label{tab:Sample}
\end{table}

\section{Line scans} 
\label{sec:LineScans}

Using the datacubes we obtain after processing, we scan for prominent lines by collapsing channels in the range of 80-400 km~$\rm s^{-1}$ (the range of expected [C~{\sc ii}] line-widths). This optimises the width of the collapsed narrowband for which the strongest point-source signal within a 1 arcsec radius of the target source (identified in the $HST$-imaging) is found, if any. Scanning through the datacube also enables us to identify any lines present from serendipitous sources.

We replicate this scanning in the sign-inverse of the datacube, to check if any noise is comparable to the signal detected from the target source and assess the robustness of any tentative ($>$3$\sigma$) line detections. We extract spectra from identified line candidates by summing all pixels detected at S/N > 2 in the collapsed narrowband. 

\subsection{A1703-zD1}
\label{sec:zD1}

\begin{figure}
    \centering
    \includegraphics[width=\columnwidth]{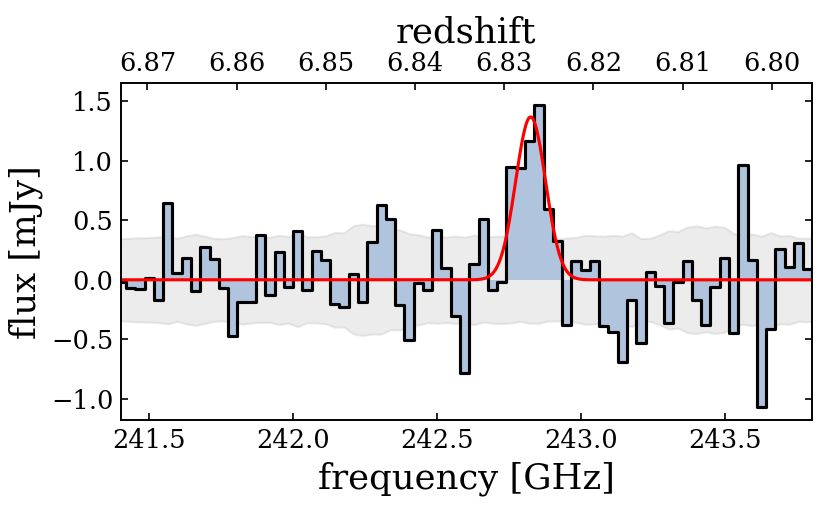}
    \caption{The spectrum of A1703-zD1 extracted by summing the flux over the source from all pixels with S/N$>$2, using the tapered imaging. The red line shows the best fit Gaussian line profile. The grey shaded region gives the measured rms throughout the spectrum.}
    \label{fig:A1703Spectra}
\end{figure}

Taking our NOEMA observations of A1703-zD1 in isolation we find a signal at 242.90$\pm$0.01 GHz with S/N=4.6, co-spatial with the lower bright clump and extended along the lensed arc, visible in $HST$-imaging (See Figure~\ref{fig:A1703Contour}). From previous observations by \cite{Schaerer15}, we independently identified a tentative detection of [C~{\sc ii}] with S/N=3.4 at 242.853$\pm$0.009 GHz. The top panel of Figure~\ref{fig:A1703Contour} presents the line detection after combining both data-sets (see Section~\ref{sec:observations}), collapsed in a 120 km~$\rm s^{-1}$ channel centered on 242.8 GHz, with a peak S/N of 5.5. To obtain an unresolved measurement of A1703-zD1 we taper the datacube to match the spatial extent of A1703-zD1 (described in Section~\ref{sec:observations}). From this we find a detection at 242.82$\pm$0.01 GHz with an increase in the peak S/N to 6.1, yielding a best-fit $z_{\rm [CII]}$ = 6.8269$\pm$0.0004 from the extracted spectrum (See Figure~\ref{fig:A1703Spectra}). This higher S/N when matching the spatial extent of the target (and alignment with the rest-frame UV emission) is a clear confirmation of the detection. We therefore use the tapered imaging for our final measurements. We collapse the datacube over the frequency range 242.72 -- 242.92 GHz and measure the total flux using 2D gaussian fitting from the routine \textsc{imfit} in the Common Astronomy Software Application (CASA; \cite{McMullin07}).

We also note that within the same datacube, we identify a serendipitous line signal at 241.59 GHz with S/N $\sim$ 14, (see spectrum in Fig~\ref{fig:A1703seren}) $\sim$ 8 arcsec offset to A1703-zD1. The signal is co-spatial with a foreground source (J131459.75+515008.6) which has $z_{\rm phot}\sim$ 0.44 $\pm$ 0.11 \citep[found in Sloan Digital Sky Survey DR12,][]{Alam15}. We therefore identify the line as CO(3-2) (rest frame wavelength 867$\rm\mu$m) at $z_{\rm spec}$ = 0.43127 $\pm$ 0.00003, consistent with the $z_{\rm phot}$. We also measure a 68$\mu$Jy continuum flux from this source with S/N of 5.1.

\subsection{EGS-5711 \& EGS-1952}
\label{sec:EGS5711}

When scanning for [C~{\sc ii}] in EGS-5711 we find a tentative signal with S/N = 4.6 at 246.16 $\pm$ 0.02 GHz, 0.5 arcsec away from the centre of the target. This line, if real, would put the source at $z$ = 6.7263 $\pm$ 0.0006. A second signal at 242.10 $\pm$ 0.03 GHz with S/N = 3.4 is also found co-spatial with EGS-5711. The corresponding spectra and contour overlay plots for the S/N = 4.6 signal can be found in the appendix; figures \ref{fig:5711tent} and \ref{fig:5711tentspectra}. In figures \ref{fig:EGS5711_LSB} and \ref{fig:EGS5711_USB} in the appendix we also present the spectra at the location of the source to illustrate the lack of detection there. Performing the same scan on the inverse datacube (searching within a radius of 15 arcsec), we find several signals with equivalent or higher S/N, the highest of which is S/N = 5.3. Given our 15 times larger search area, we expect $\sim$1 spurious signal above 4.5$\sigma$ within 1 arcsec radius of our target. More observations are therefore required to confirm any detection. 

For EGS-1952 we find no evidence for [C~{\sc ii}] with S/N>3, co-spatial with the source. Spectra at the location of the source can be found in the appendix (Figures~\ref{fig:EGS1952_LSB}, \ref{fig:EGS1952_USB}.) Further, we find no continuum signal for both EGS-5711 and EGS-1952 and as a result, we only provide upper limits for the continuum flux and an upper limit estimate for the [C~{\sc ii}] line flux, based on the spectral line obtained for A1703-zD1, scaled to the median rms of the EGS-5711 and EGS-1952 datacubes.

We can either interpret these non-detections as faint [C~{\sc ii}] lines due to the low dust content of these sources or alternatively it is possible that the limited frequency range that was scanned for each object missed the line emission entirely. Deeper and wider scans would be needed to distinguish between these two interpretations (see Appendix~\ref{app:A}).

\begingroup
\renewcommand{\arraystretch}{1.25}
\begin{table}
\begin{tabular}{|c|c|c|c|}
\hline
Target ID & A1703-zD1 & EGS-5711 & EGS-1952\\ 
\hline    
$z_{\rm [CII]}$ & 6.8269$\pm$0.0004 & -- & -- \\
S/N & 6.1 & -- & -- \\
$[\rm C~{II}]$ flux (Jy km~$\rm s^{-1}$) & 0.28$\pm$0.06 & < 0.17$^a$ & < 0.21$^a$ \\
$\rm FWHM_{[CII]}$ (km~$\rm s^{-1}$) & 155$\pm$31 & -- & -- \\
Continuum flux ($\rm \mu$Jy) & < 24.5$^a$ & < 27.4$^a$ & < 28.7$^a$\\
$L_{\rm[CII]}$ (10$^8$ $L_\odot$) & 0.35$\pm$0.07 ($_{-0.04}^{+0.29}$)$^c$ & < 1.8$^a$ & < 2.3$^a$ \\
SFR$_{\rm[CII]}$ ($M_\odot$ yr$^{-1}$) & 4.3$\pm$0.9 ($_{-0.04}^{+0.29}$)$^c$ & < 23.9$^a$ & < 29.5$^a$ \\
$L_{\rm UV}$ (10$^{11}$ $L_\odot$) & 0.39$\pm$0.04 ($_{-0.04}^{+0.41}$)$^c$ & 1.2$\pm$0.1 & 1.0$\pm$0.1 \\
SFR$_{\rm UV}$ ($M_\odot$ yr$^{-1}$) & 6.7$\pm$0.6 ($_{-0.61}^{+6.41}$)$^c$ & 20.2$\pm$1.9 & 17.9$\pm$1.7 \\
$L_{\rm IR}$ (10$^{11}$ $L_\odot$) & < 0.35 (< 0.68)$^{a,b,c}$ & < 3.6$^{a,b}$ & < 3.9$^{a,b}$ \\
SFR$_{\rm IR}$ ($M_\odot$ yr$^{-1}$) & < 5.2 (< 10.1)$^{a,b,c}$ & < 53.2$^{a,b}$ & < 57.7$^{a,b}$\\
log(IRX) & < $-$0.05$^a$ & < 0.48$^a$ & < 0.57$^a$ \\
\hline
\end{tabular}
\caption{Galaxy Properties $^a$3$\sigma$ upper limits, $^b$assuming a grey body approximation with $T$ = 50K and $\beta$ = 1.5, $^c$corrected for a magnification of $\mu$=9.0 (the uncertainty due to the lensing magnification are also shown in brackets). Observed quantities are uncorrected for magnification, while the derived physical properties use the magnification correction.}
\label{tab:GalProperties}
\end{table}
\endgroup

\section{Source properties}
\label{sec:SourceProperties}

Here we present the physical properties of our three targets derived from our observations and comparisons to the literature. A full table of galaxy properties is shown in Table~\ref{tab:GalProperties}.

\subsection{$L_{\rm [CII]}$ vs SFR relation}
\label{sec:CII Relation}

\begin{figure}
\centering
    \includegraphics[width=\columnwidth]{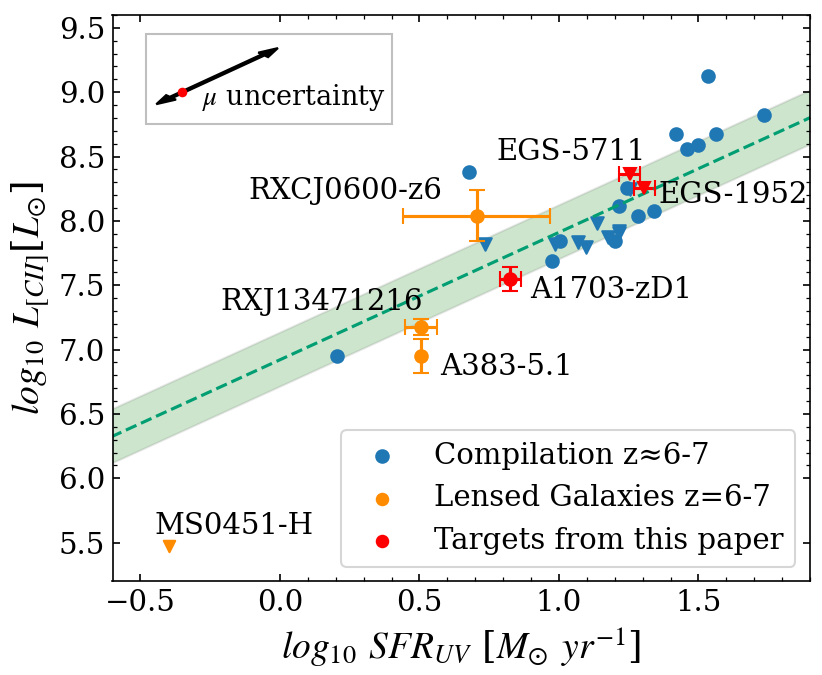}
    \caption{The [C~{\sc ii}] line luminosity as a function of the star-formation rate, derived from the ultraviolet luminosity. The locally observed \citealt{DeLooze14} relation is indicated by the dotted green line with the errors shown by the green shaded region. Sources with 3$\sigma$ upper limits are shown by downward triangles. Our three targets are presented as red points. The black arrows in the top left indicate the magnification uncertainty for A1703-zD1. Changes in magnification affect both $L_{\rm [CII]}$ and SFR$_{\rm UV}$, such that A1703-zD1 moves along the local relation. We use a compilation of $z$ $\approx$ 6 -- 7 sources for comparison \citep[compiled from][]{Matthee19}, shown in blue. In orange we highlight four lensed sources from the literature \citep{Knudsen16,Bradac17, Fujimoto21} for which three show an apparent offset from the local $L_{\rm[CII]}$ -- SFR relation, one of which shows a very clear offset (MS0451-H, discussed more in Section~\ref{sec:CII Relation}). }
    \label{fig:LitCompare}
\end{figure}

In Figure~\ref{fig:LitCompare} we present the measured [C~{\sc ii}] luminosity as a function of SFR$_{\rm UV}$ and include sources from the literature for comparison \citep{Matthee19,Knudsen16,Bradac17, Fujimoto21}. The UV SFRs are calculated using the  \cite{Kennicutt12} conversion, using a Kroupa IMF.
EGS-5711 and EGS-1952 are given as 3$\sigma$ upper limits.

In the local universe we see a tight $L_{\rm [CII]}$ -- SFR relationship as shown by the \citealt{DeLooze14} relation ($\log{\rm SFR}$ = -- 6.99 + 1.01 $\times$ $\log{\rm [CII]}$) in Figure~\ref{fig:LitCompare}. Recent studies of `normal' (i.e. main sequence) star-forming galaxies with redshifts at 4.4 < z < 5.9 and galaxies at z$\sim$6.5 with high Ly$\alpha$ luminositites have also shown consistency with the local relation when including the dust-obscured SFR \citep[e.g.,][]{Matthee19,Schaerer20}. In contrast, a few studies have found that lensed galaxies with a lower SFR are more likely to be below the locally observed relation \citep[e.g.,][]{Knudsen16,Bradac17}. In particularly the strongest lensed object, MS0451-H \citep{Knudsen16}, with the lowest intrinsic SFR shows the strongest deficit in $L_{\rm [CII]}$. If confirmed, this could suggest differing ISM properties in faint, and possibly more metal-poor, high-redshift galaxies. A1703-zD1 is a strongly lensed galaxy with a modest intrinsic SFR$_{\rm UV}$ of 6.7 $\pm$ 0.6 M$_\odot$ yr$^{-1}$, but we find no evidence of a significant offset to the local relation, as A1703-zD1 lies slightly below, but still within 1$\sigma$ of, the local $L_{\rm [CII]}$ -- SFR relation. We also note that including the magnification uncertainties in the calculations of $L_{\rm [CII]}$ and SFR$_{\rm [CII]}$ has no impact on the offset to the local relation as indicated by the arrows in Figure~\ref{fig:LitCompare}. However, including the 3$\sigma$ upper-limit on the SFR$_{\rm IR}$ would place A1703-zD1 below the local relation.

An important consideration in assessing a possible [C~{\sc ii}] deficit is the potential for extended [C~{\sc ii}] emission \citep[e.g.,][]{Fujimoto19,Fujimoto20} compared to the rest-frame UV data, in particular for sources with smearing due to strong gravitational lensing.
This effect was recently studied by \cite{Carniani20}, who find that [C~{\sc ii}] emission can be two times more extended than [O~{\sc iii}].  Specifically, \cite{Carniani20} show that MS0451-H \citep{Knudsen16}, nominally a 2.5$\sigma$ deviation from the local De Looze relation (Figure~\ref{fig:LitCompare}), misses $\sim$ 60 -- 80 per cent of the [C~{\sc ii}] emission if the effects of lensing are ignored. Taking into account possible extended [C~{\sc ii}] moves MS0451-H to within 1$\sigma$ of the local relation.

In A1703-zD1, we aim to account for lens smearing by tapering the datacube to match the spatial extent of A1703-zD1 in the $HST$ imaging, allowing us to obtain a more accurate flux measurement. Without tapering and using the same routine as described in Section~\ref{sec:zD1}, we calculate a [C~{\sc ii}] integrated flux value which is 51 per cent of our fiducial measurement. However, extended [C~{\sc ii}] beyond the rest-frame UV could still be missed even with our current tapering strategy, consistent with \cite{Carniani20}.

\subsection{$L_{\rm IR}$ and SFR$_{\rm IR}$ constraints}
\label{sec:IR}

As discussed in Section~\ref{sec:observations}, the continuum flux remains undetected for all our sources. We calculate upper limits for $L_{\rm IR}$ and SFR$_{\rm IR}$ by assuming an optically-thin grey-body infrared SED \citep{Casey12} using $T$ = 50K and $\beta$ = 1.5, and present the results in Table~\ref{tab:GalProperties}. We estimate that these sources are likely below the classification of Luminous Infrared Galaxies (LIRG; $L_{\rm IR}>10^{11}\,L_\odot$). We furthermore find obscured SFR below 5-58 $M_\odot\,\rm yr^{-1}$ (3$\sigma$ limits), which suggests less than 44 -- 74\% of star-formation comes out in the IR, consistent with recent studies \citep[e.g.,][]{Bowler18, Schouws21}.

We derive UV slopes ($\beta$) and upper limits on the Infrared Excess (IRX = $L_{\rm IR}$/$L_{\rm UV}$) for all our targets (see Tables~\ref{tab:Sample} and~\ref{tab:GalProperties}) and find that the upper-limits for EGS-5711 and EGS-1952 are consistent with either the \cite{Meurer99} relation or a SMC-like dust attenuation.

However, given a moderately red UV slope for A1703-zD1 of $\beta$ $\sim$ -1.56, we find this galaxy to be more consistent with an SMC like dust attenuation law, in agreement with stacking results of faint LBGs at $z$ $\approx$ 2 -- 10 \citep[][ though see \citealt{Schouws21} for a discussion on the impact of the assumed dust temperature]{Bouwens16, Fudamoto20}.

\subsection{Velocity Structure}

\begin{figure}
\centering
    \includegraphics[width=\columnwidth]{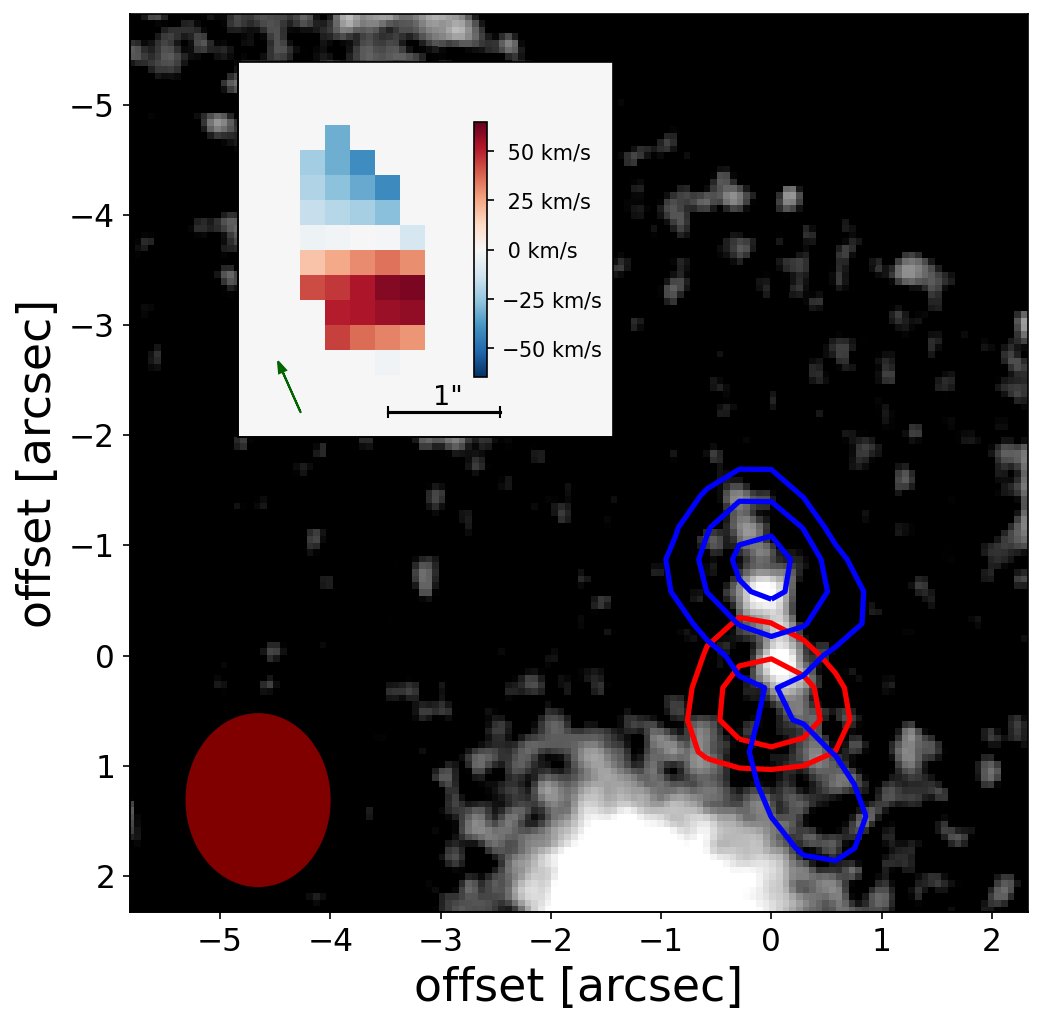}
    \caption{ \textbf{Main Panel:} HST imaging overlaid with contours from the untapered datacube corresponding to 2,3 and 4$\sigma$ [C~{\sc ii}] emission collapsed over channels redwards (red contours) and bluewards (blue contours) of the measured line centre, showing evidence for a velocity gradient. The observations are spatially resolved, as shown by the filled ellipse indicating the beam size (1.57 $\times$ 1.31 arcsec).
    \textbf{Inset Panel:} First moment map of the detected [C~{\sc ii}] from A1703-zD1 (showing pixels above 2$\sigma$). We measure a velocity difference over the galaxy of 103 $\pm$ 22 km~$\rm s^{-1}$. We show the green arrow to illustrate the stretching of the arc in the image plane with respect to the source-plane image of this source. }
    \label{fig:VelocityGrad}
\end{figure}

We use the spatial extent of the [C~{\sc ii}] detection to investigate the velocity structure of A1703-zD1. We see a velocity gradient across A1703-zD1 shown in Figure~\ref{fig:VelocityGrad}. We find a maximum projected velocity difference over the galaxy ($\Delta v_{\rm obs}$) of 103 $\pm$ 22 km~$\rm s^{-1}$ from the first moment map. Such a velocity gradient could be the signature of a rotating disk, whilst another possibility is a merger of [C~{\sc ii}] emitting galaxies. Similar velocity gradients are present in previous observations of high redshift galaxies \citep[e.g.,][]{Smit18,Hashimoto19,Matthee19,Fujimoto21} as well as in simulations \citep{Dekel14}. 

To determine the likelihood of a disk-like rotation we compare the projected velocity range of a galaxy with the velocity dispersion of the system using $\Delta v_{\rm obs}$/2$\sigma_{\rm tot}$, where a ratio of >0.4 indicates a likely rotation dominated system \citep{Schreiber09}. We find $\Delta v_{\rm obs}$/2$\sigma_{\rm tot}$ = 0.79 $\pm$ 0.23 for A1703-zD1, which supports the interpretation of a possibly rotation dominated system. \cite{Bradley12} find three distinct star-forming clumps with an extended linear morphology in the source-plane reconstruction of A1703-zD1. In Figure~\ref{fig:VelocityGrad} we show the deflection due to the lensing magnification, using the LTM strong-lensing model published by \cite{Zitrin10}. This stretching of the source-plane leads to an effective increase in resolution of 3.5$\times$, in the direction of the green arrow shown in Figure~\ref{fig:VelocityGrad}. The red and blue components identified in Figure~\ref{fig:VelocityGrad} appear co-spatial with the two brightest clumps from \cite{Bradley12}. Clumps like this have been identified previously in high redshift galaxies and can be attributed to merging galaxies or large clumpy star formation due to the increased gas content in high-redshift galaxies. Higher resolution [C~{\sc ii}] observations will be required to confirm a disk-like structure in A1703-zD1.

If we assume ordered circular rotation we can use the FWHM of [C~{\sc ii}] to estimate the dynamical mass. We use $M_{\rm dyn}$ = $1.16$ x $10^5V_{\rm cir}^2D$ \citep[][]{Capak15}, where $V_{\rm cir}$ = 1.763 x $\sigma_{\rm [CII]}$ / $\sin{i}$  in km~$\rm s^{-1}$, $i$ is the disk inclination angle, and $D$ is the disk diameter in kpc (we use $D=4$ kpc found in \citealt{Bradley12}; measured in the reconstructed sources-plane image). Assuming a viewing angle of 45 degrees we find $M_{\rm dyn}$ = 12.1 $\pm$ 4.8 x $10^9$ $M_\odot$. The stellar mass of A1703-zD1 was esimated to be 0.7 $\pm$ 0.1 x $10^9$ $M_\odot$ \citep[][]{Bradley12}. This stellar mass is only $\sim$ 6 per cent of the total dynamical mass that we measure. This is somewhat low compared to dynamical measurements from H$\alpha$ surveys at cosmic noon; for example, \cite{Stott16} find a value of 22 $\pm$ 11 per cent in a sample of 584 z $\sim$ 1 galaxies and \cite{Wuyts16} find a value of $32_{-7}^{+8}$ per cent in star forming galaxies 0.6 < z < 2.6. Larger and more detailed samples of dynamical measurements at $z\sim7$ will be needed to establish evolutionary trends with redshift. We also note that there are large uncertainties on the dynamical mass, as there is a large dependency on the viewing angle (which is unknown). However, this low ratio does suggest that A1703-zD1 is a gas rich system, similar to typical star-forming galaxies 3 Gyr later in cosmic time.



\section{Summary and conclusions}
\label{sec:Conclusions}
We have presented new NOEMA observations, scanning for the [C~{\sc ii}]158$\mu$m line in three Lyman break galaxies with photometric redshifts at $z=6.6 - 6.9$. Our main findings are as follows:
\begin{itemize}
    \item We detect [C~{\sc ii}] in one of our three sources, confirming the redshift at $z$ = 6.8269 $\pm$ 0.0004 for the strongly lensed galaxy A1703-zD1 (6.1$\sigma$). Our non-detections are consistent with these being dust-poor galaxies with low [C~{\sc ii}] luminosity. 
    \item We carefully account for any extended emission of [C~{\sc ii}] due to lens smearing in A1703-zD1 and find the [C~{\sc ii}] luminosity to be consistent with, but slightly below, the local L$_{\rm [CII]}$ -- SFR relation. 
    \item No continuum emission is detected in any of the three targeted sources, suggesting less than 44 -- 74\% of star-formation comes out in the IR. For A1703-zD1, our results are most consistent with an SMC attenuation curve. 
    \item We see a velocity gradient across A1703-zD1, with a kinematic ratio that suggests a possible rotation dominated system, though higher resolution [C~{\sc ii}] observations will be needed to confirm this.
\end{itemize}

Over the last few years ALMA has demonstrated its role as a ``redshift machine'' in the Epoch of Reionisation by confirming galaxies out to redshift $z$ = 9. In this paper we have demonstrated the ability of NOEMA to search for [C~{\sc ii}] in `normal' star-forming galaxies at $z$ > 6, complementing ALMA by observing EoR galaxies in the Northern Hemisphere, with [C~{\sc ii}] as a reliable spectroscopic tracer of these distant systems. With the launch of JWST this capability will be particularly useful for rare, lensed sources and intrinsically luminous objects that will be discovered far outside the limited JWST survey area using the next generation of large area surveys, such as the Euclid mission and the Rubin observatory.

\section*{Acknowledgements}
We thank the anonymous referee for their constructive feedback and useful suggestions. RS acknowledges support from a STFC Ernest Rutherford Fellowship (ST/S004831/1). AZ acknowledges support by Grant No. 2020750 from the United States-Israel Binational Science Foundation (BSF) and Grant No. 2109066 from the United States National Science Foundation (NSF), and by the Ministry of Science \& Technology, Israel. JH gratefully acknowledges support of the VIDI research program with project number 639.042.611, which is (partly) financed by the Netherlands Organisation for Scientific Research (NWO).

\section*{Data availability}
The reduced data underlying this article will be shared on reasonable request to the corresponding author.




\bibliography{refs}{}
\bibliographystyle{mnras}



\appendix

\section{Photometric redshift coverage}
\label{app:A}

Here we present the SEDs for all three targets in this paper, alongside the photometric redshift probability distributions, redshift coverage of the observations and in the case of A1703-zD1, the measured $z_{\rm [CII]}$. The SED fitting was done using the software Easy and Accurate Zphot from Yale (EAZY; \cite{Brammer08}), see \cite{Smit15} for a full description.

We note that for EGS-5711, NOEMA did not cover the central $z_{\rm phot}$ of $6.47_{-0.10}^{+0.11}$. The systematic uncertainties in estimating photometric redshifts in extreme emission line galaxies are discussed in section 2.1, which motivated a line search at $z\sim6.6-6.9$ based on the \citet{Smit15} colour selection.

\begin{figure}
    \centering
    \includegraphics[width=0.5\textwidth]{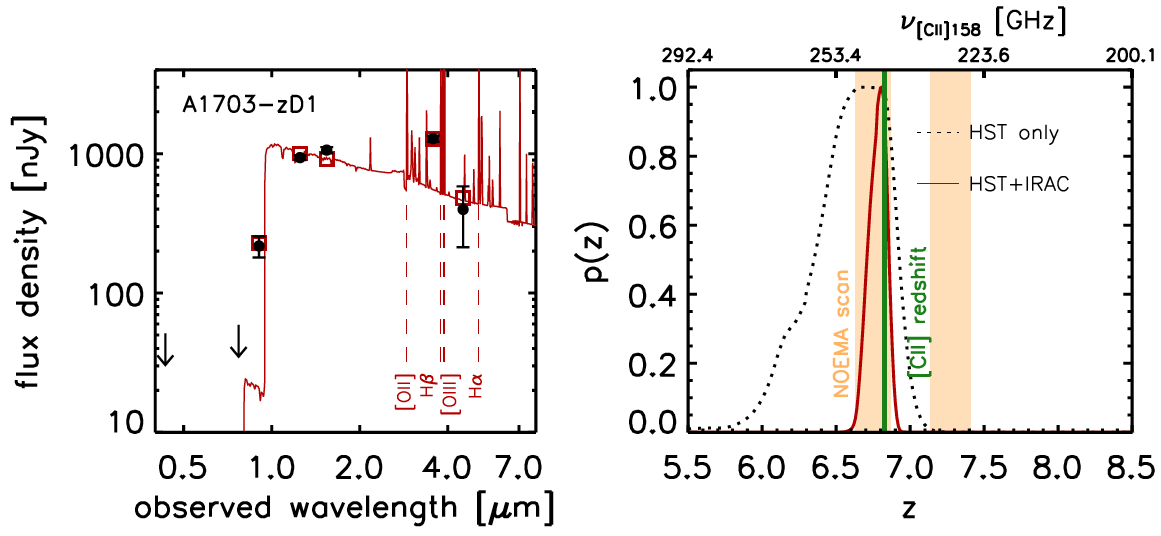}
    \caption{A1703-zD1 SED (\textbf{left panel}) and redshift probability distribution distribution (\textbf{right panel}). Coverage by the NOEMA sidebands (USB and LSB) is indicated by the shaded orange region. The dotted black line and solid red line represent the probability for the target to be at a certain redshift based on HST only and HST+IRAC data, respectively. Here the USB was selected to cover the peak probability in the photometric redshift range, which is indeed where the detection of [C~{\sc ii}] was made, indicated by the solid green line.}
    \label{fig:A1703_SED}
\end{figure}

\begin{figure}
    \centering
    \includegraphics[width=0.5\textwidth]{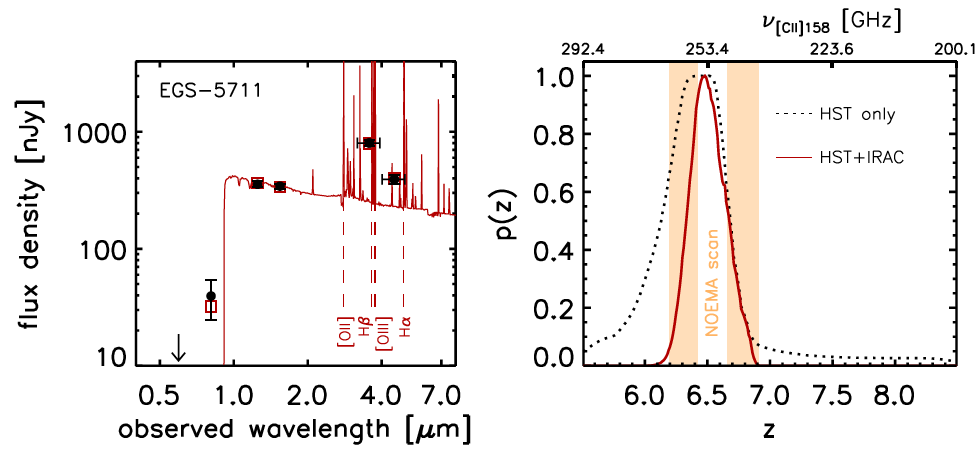}
    \caption{EGS-5711 SED (\textbf{left panel}) and redshift probability distribution distribution (\textbf{right panel}). Coverage by the NOEMA sidebands (USB and LSB) is indicated by the shaded orange region. The dotted black line and solid red line represent the probability for the target to be at a certain redshift based on HST only and HST+IRAC data, respectively. This target was selected to be at $z\sim6.6-6.9$ based on HST+IRAC data colour criteria \citep{Smit15}, however, the peak of the p(z) suggests a possible lower redshift. A broader and deeper line scan will be needed to establish the spectroscopic redshift of this source.}
    \label{fig:EGS5711_SED}
\end{figure}

\begin{figure}
    \centering
    \includegraphics[width=0.5\textwidth]{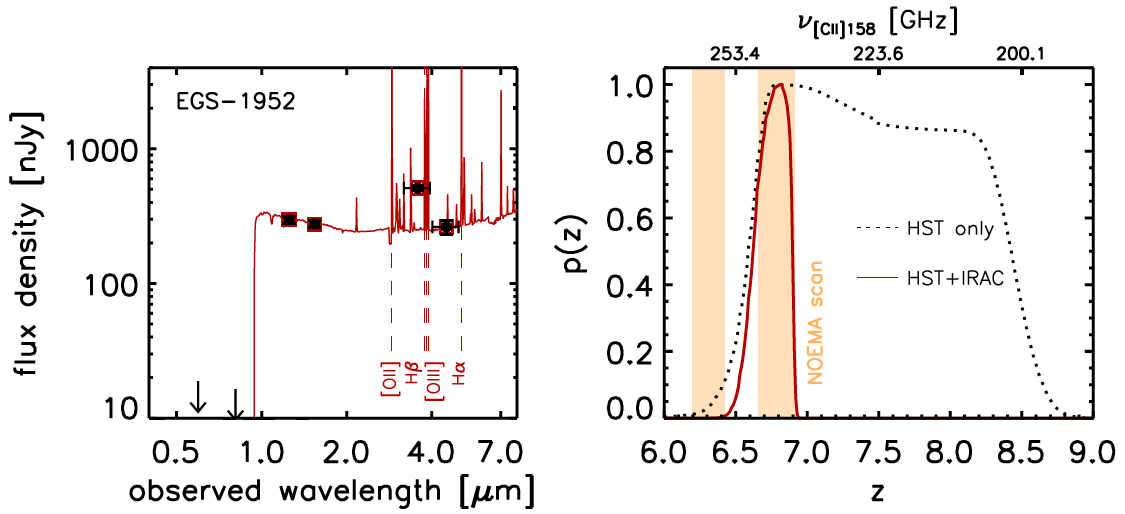}
    \caption{EGS-1952 SED (\textbf{left panel}) and redshift probability distribution (\textbf{right panel}). Coverage by the NOEMA sidebands (USB and LSB) is indicated by the shaded orange region. The dotted black line and solid red line represent the probability for the target to be at a certain redshift based on HST only and HST+IRAC data, respectively.}
    \label{fig:EGS1952_SED}
\end{figure}

\section{EGS-5711 tentative detection}

\begin{figure}
    \centering
    \includegraphics[width=0.5\textwidth]{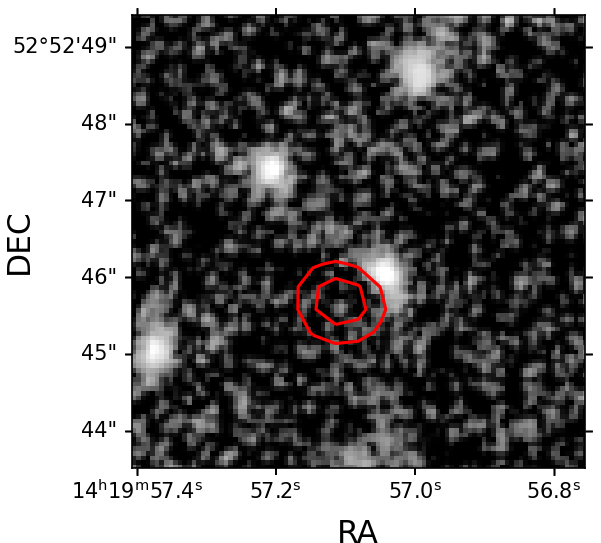}
    \caption{The S/N=4.6 signal found for EGS-5711. Background shows the HST $H$-band image with red contours corresponding to 3 and 4 $\sigma$ [C~{\sc ii}] emission. The extracted spectrum is also shown in Figure~\ref{fig:5711tentspectra}. This source would require further observations to confirm a robust detection.}
    \label{fig:5711tent}
\end{figure}

\begin{figure}
    \centering
    \includegraphics[width=0.5\textwidth]{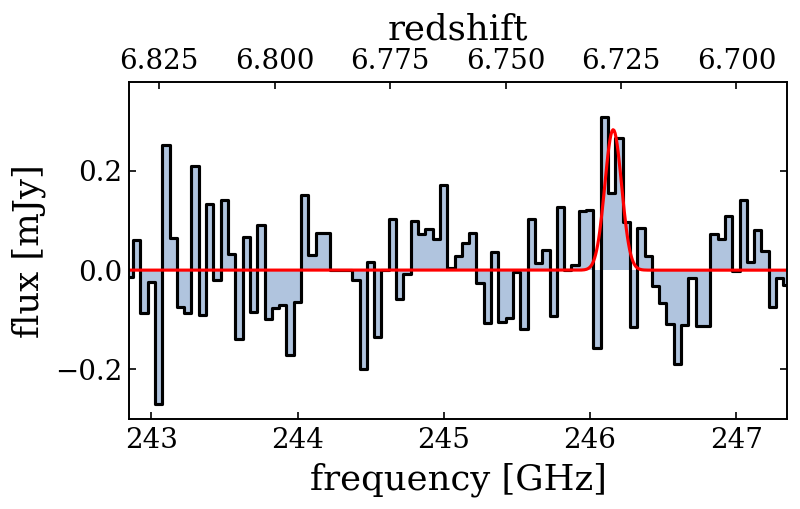}
    \caption{EGS-5711 spectrum of the S/N = 4.6 signal, extracted from the region of the contours shown in Figure~\ref{fig:5711tent}, with the best-fit Gaussian line profile in red.}
    \label{fig:5711tentspectra}
\end{figure}

Here we present a 4.6$\sigma$ signal for EGS-5711, with the contours overlayed on the HST imaging in Figure~\ref{fig:5711tent} and the corresponding spectra in Figure~\ref{fig:5711tentspectra}. The spectral extraction is the same as described in Section~\ref{sec:LineScans}. The weak signal is very narrow and clearly offset from the HST target, which likely makes this a spurious detection. 

\section{Empty scans}

\begin{figure}
    \centering
    \includegraphics[width=0.5\textwidth]{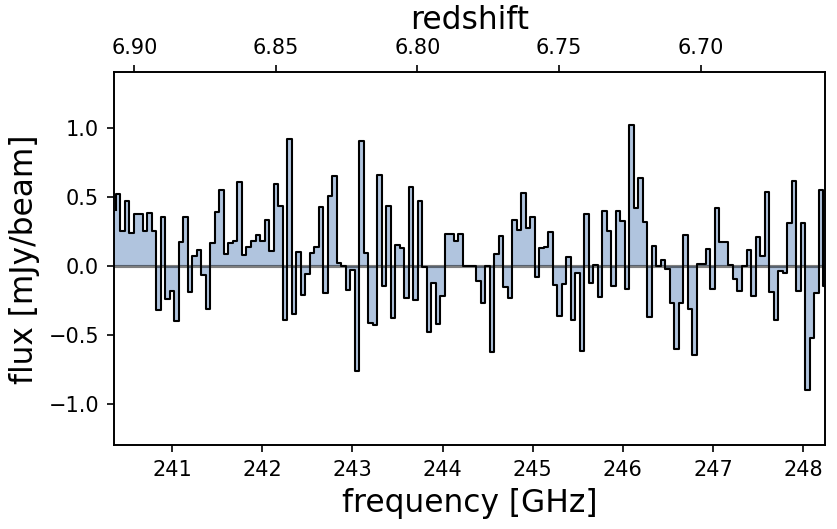}
    \caption{LSB spectra taken at the location of EGS-5711. We find no robust detection and present upper limits in table \ref{tab:GalProperties}.}
    \label{fig:EGS5711_LSB}
\end{figure}

\begin{figure}
    \centering
    \includegraphics[width=0.5\textwidth]{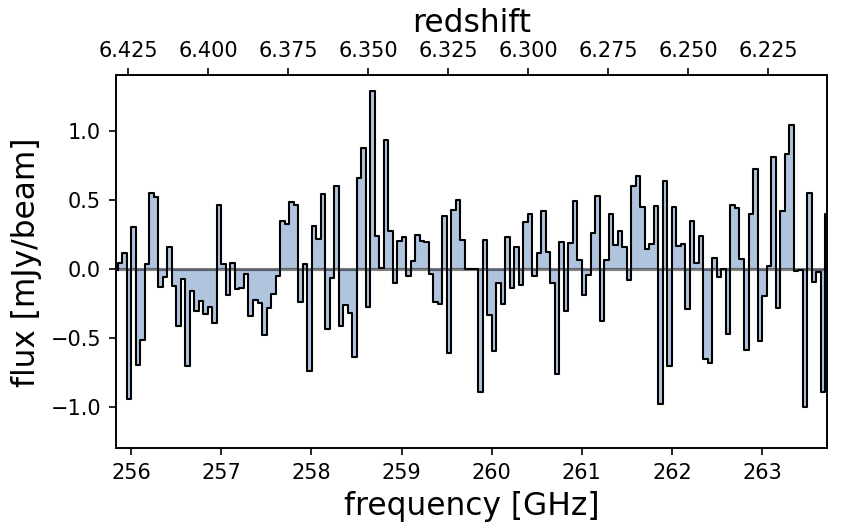}
    \caption{USB spectra taken at the location of EGS-5711. }
    \label{fig:EGS5711_USB}
\end{figure}

\begin{figure}
    \centering
    \includegraphics[width=0.5\textwidth]{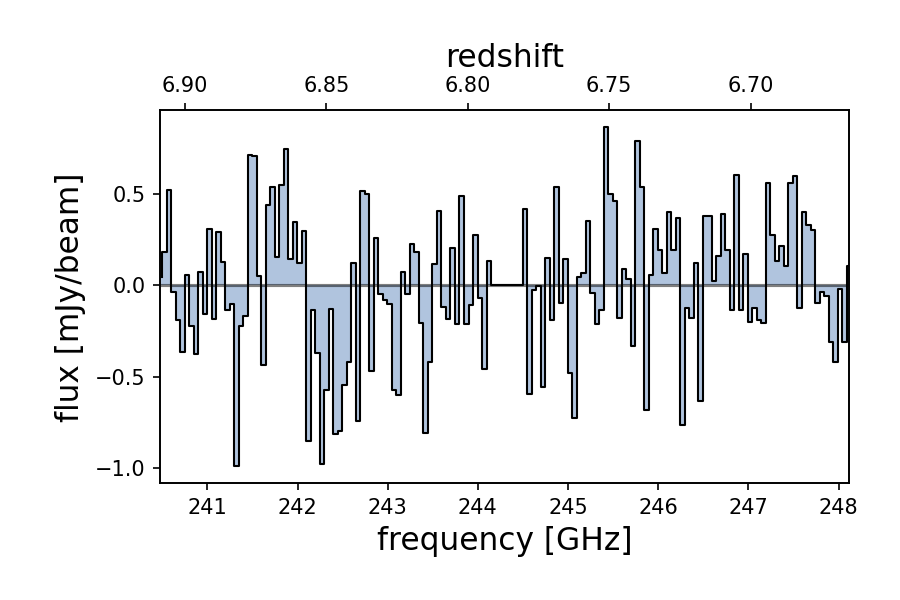}
    \caption{LSB spectra taken at the location of EGS-1952. We find no detection and present upper limits in table \ref{tab:GalProperties}.}
    \label{fig:EGS1952_LSB}
\end{figure}

\begin{figure}
    \centering
    \includegraphics[width=0.5\textwidth]{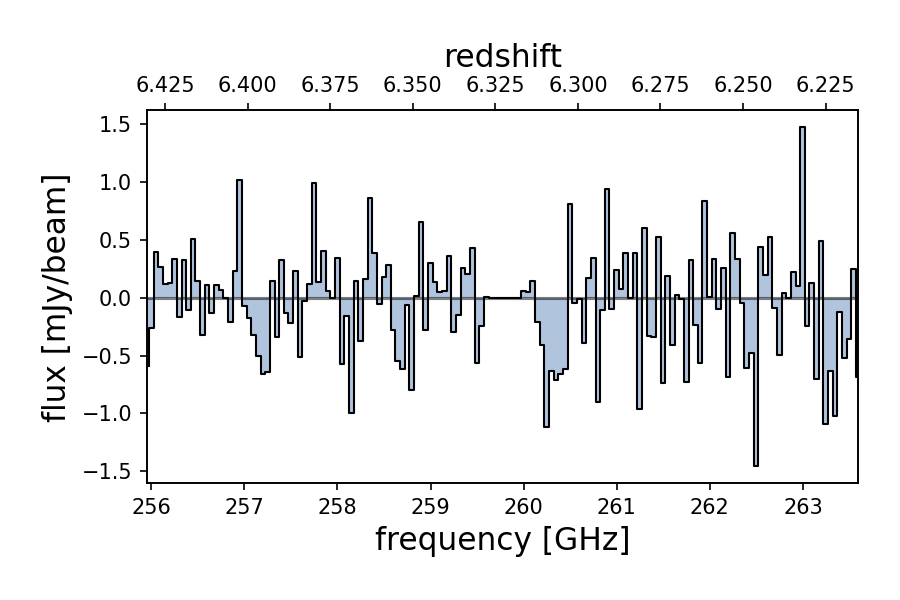}
    \caption{USB spectra taken at the location of EGS-1952. }
    \label{fig:EGS1952_USB}
\end{figure}

In Figures \ref{fig:EGS5711_LSB}, \ref{fig:EGS5711_USB}, \ref{fig:EGS1952_LSB} and \ref{fig:EGS1952_USB} we present the empty line scans in the LSB and USB for both EGS targets.  The extractions were taken simply from the central pixel of the observations. 

\section{Serendipitous source in a1703-zD1 cube}

\begin{figure}
    \centering
    \includegraphics[width=0.5\textwidth]{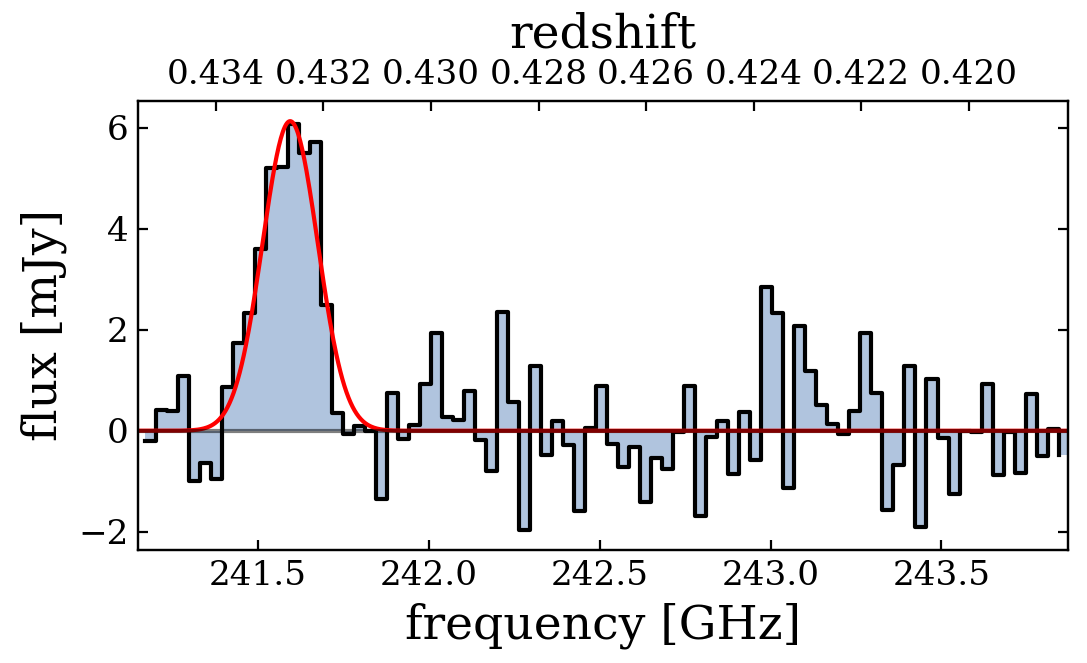}
    \caption{Spectrum of the serendipitous line identified in the datacube of A1703-zD1, with the best-fit Gaussian line profile in red. The redshifts plotted along the upper horizontal axis are based on the assumption the line detection is CO(3-2) for the source J131459.75+515008.6. The S/N of the line is 14.}
    \label{fig:A1703seren}
\end{figure}

\begin{figure}
    \centering
    \includegraphics[width=0.5\textwidth]{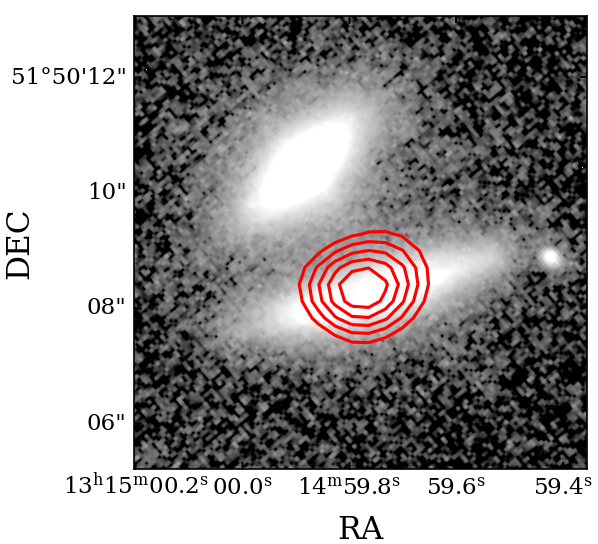}
    \caption{Serendipitous source (J131459.75+515008.6) with red contours overlayed corresponding to 4, 6, 8, 10 and 12 $\sigma$, identified as CO(3-2) emission. Background shows the HST $z_{850}$ imaging of Abell 1703. }
    \label{fig:A1703serenim}
\end{figure}

Here we present a serendipitous 14$\sigma$ detection in the datacube of A1703-zD1, with the spectra shown in \ref{fig:A1703seren} and the line contours overlayed on the HST imaging in \ref{fig:A1703serenim}. The spectral extraction is the same as described in Section~\ref{sec:LineScans}. The detection is co-spatial with a foreground source (J131459.75+515008.6) which has $z_{\rm phot}\sim$0.44 $\pm$ 0.11 \citep[found in Sloan Digital Sky Survey DR12,][]{Alam15}. We therefore attribute the line detection to CO(3-2) (rest frame wavelength 867$\rm\mu$m) putting the source at $z_{\rm spec}$ = 0.43127 $\pm$ 0.00003, consistent with the $z_{\rm phot}$.

\bsp	
\label{lastpage}
\end{document}